\documentclass[Physsubmission, Phys]{SciPost}

\hypersetup{
    colorlinks,
    linkcolor={red!50!black},
    citecolor={blue!50!black},
    urlcolor={blue!80!black}
}

\usepackage[bitstream-charter]{mathdesign}
\DeclareSymbolFont{usualmathcal}{OMS}{cmsy}{m}{n}
\DeclareSymbolFontAlphabet{\mathcal}{usualmathcal}

\begin{document}

\begin{center}{\Large \textbf{
Recent results from the TOTEM collaboration and the discovery of the odderon\\
}}\end{center}

\begin{center}
C. Royon\textsuperscript{1},
\end{center}

\begin{center}
{\bf 1} Department of Physics and Astronomy, The University of Kansas, Lawrence, USA
\\
* christophe.royon@cern.ch
\end{center}

\begin{center}
\today
\end{center}


\definecolor{palegray}{gray}{0.95}
\begin{center}
\colorbox{palegray}{
  \begin{tabular}{rr}
  \begin{minipage}{0.1\textwidth}
    \includegraphics[width=30mm]{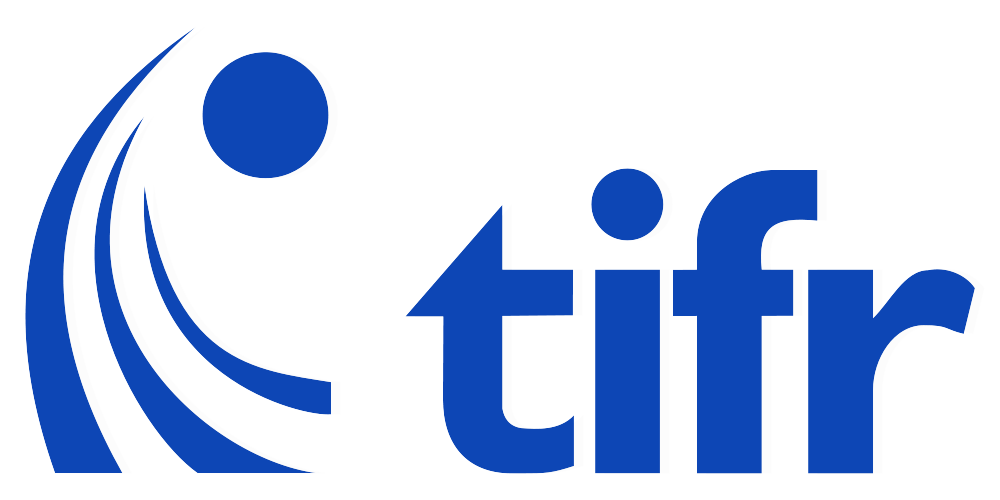}
  \end{minipage}
  &
  \begin{minipage}{0.85\textwidth}
    \begin{center}
    {\it 21st International Symposium on Very High Energy Cosmic Ray Interactions (ISVHE- CRI 2022)}\\
    {\it Online, 23-27 May 2022} \\
    \doi{10.21468/SciPostPhysProc.?}\\
    \end{center}
  \end{minipage}
\end{tabular}
}
\end{center}

\section*{Abstract}
{\bf
We describe the most recent results from the TOTEM collaboration on elastic, inelastic and total cross sections as well as the odderon discovery by the D0 and TOTEM collaborations.
}

\vspace{10pt}
\noindent\rule{\textwidth}{1pt}
\tableofcontents\thispagestyle{fancy}
\noindent\rule{\textwidth}{1pt}
\vspace{10pt}

In this short report, we will first discuss the most recent measurements from the TOTEM collaboration of elastic, inelastic and total cross sections at different center-of-mass energies of the LHC, namely 2.76, 7, 8 and 13 TeV. In a second part of this report, we will discuss the odderon discovery obtained by the D0 and TOTEM collaborations by comparing $pp$ and $p \bar{p}$ elastic interactions.

\section{Measurements of the elastic, inelastic and total cross sections by the TOTEM collaboration}

To study elastic collisions at the LHC between protons, one detects and measures the intact protons  in dedicated detectors called roman pots. Each proton is scattered at some angles and can lose or gain some momentum. The fact that the protons are intact is due to the fact that a colorless object (at least two gluons) is exchanged between the two protons.
The advantage of the TOTEM detector is that there are two inelastic telescopes called $T1$ and $T2$ that were installed in the very forward region covering the domain in rapidity of $3.1 < |\eta|<4.7$ and $5.3 < |\eta| < 6.5$ allowing to veto on any particle produced in the very forward region, or to select cleanly elastic events. In addition, it is possible to run the LHC for different center-of-mass energies (2.76, 7, 8 and 13 TeV), with different values of $\beta^*$ allowing to increase the domain of measurement in $|t|$, the 4-momentum transferred square at the proton vertex.

The TOTEM collaboration developed different methods to measure the total cross section $\sigma_{tot}$ based on the optical theorem
\begin{eqnarray}
L \sigma_{tot}^2 &=& \frac{16 \pi}{1 + \rho^2} (dN_{el}/dt)_{t=0}  \\
L \sigma_{tot} &=& N_{el} + N_{inel} 
\end{eqnarray}
where $L$ is the luminosity, $\rho= \frac{Re(A^N(0))}{Im (A^N(0))} $, the ratio of the real and the imaginary part of the nuclear amplitude at $t=0$, and $N_{el}$ and $N_{inel}$ the number of elastic and inelastic events.  Three different methods can thus be defined to measure the total cross section $\sigma_{tot}$, namely
the luminosity independent measurement
\begin{eqnarray}
\sigma_{tot} = \frac{16 \pi}{(1 + \rho^2)}
\frac{(dN_{el}/dt)_{t=0}}{(N_{el}+N_{inel})} 
\end{eqnarray}
the luminosity dependent measurement
\begin{eqnarray}
\sigma_{tot}^2 = \frac{16 \pi}{(1 + \rho^2)} \frac{1}{L} (dN_{el}/dt)_{t=0}
\end{eqnarray}
and the $\rho$ independent measurement
\begin{eqnarray}
\sigma_{tot} = \sigma_{el} + \sigma_{inel}.
\end{eqnarray}
All methods lead to consistent measurement of $\sigma_{tot}$ at different center-of-mass energies~\cite{totem1,totem3,totem4,totem5,totem6}.

\begin{figure}[h]
\centering
\includegraphics[width=0.7\textwidth]{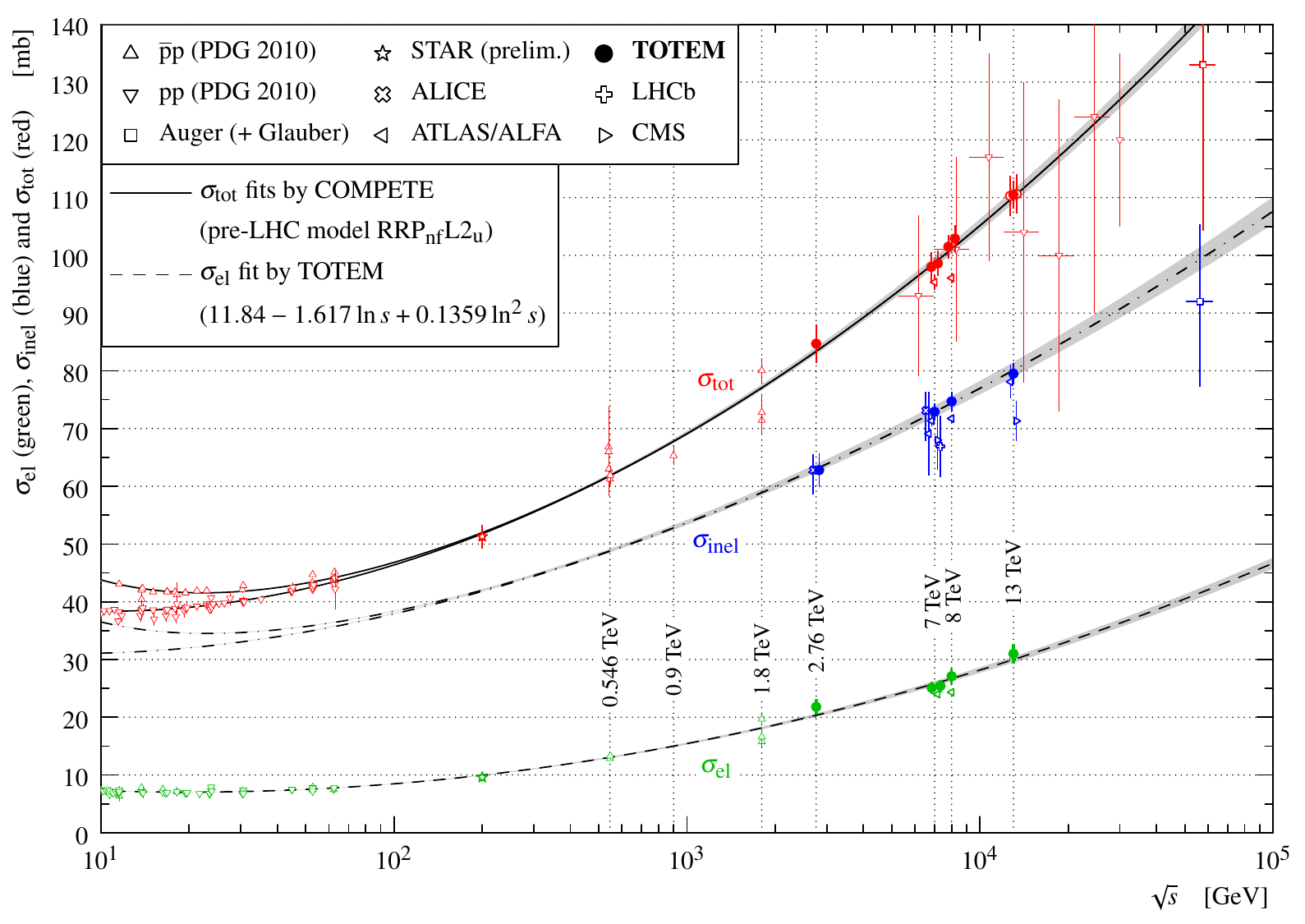}
\caption{Elastic, inelastic and total cross section measurements as a function of the center-of-mass energy $\sqrt{s}$.}
\label{elastic}
\end{figure}

The measurements by the TOTEM collaboration of the elastic, inelastic and total cross sections are displayed in Fig.~\ref{elastic}. There is a good agreement with the cosmic ray measurements at high energies (within uncertainties) and a discrepancy of about 1.9$\sigma$ has been found with ATLAS~\cite{atlas2} at 8TeV~\footnote{A recent measurement of the total cross section from ATLAS at 13 TeV was recently released at the ICHEP 2022 conference after the ISVHECRI workshop  and shows the same level of discrepancy between ATLAS and TOTEM~\cite{atlas1}.}.

\begin{figure}[h]
\centering
\includegraphics[width=0.7\textwidth]{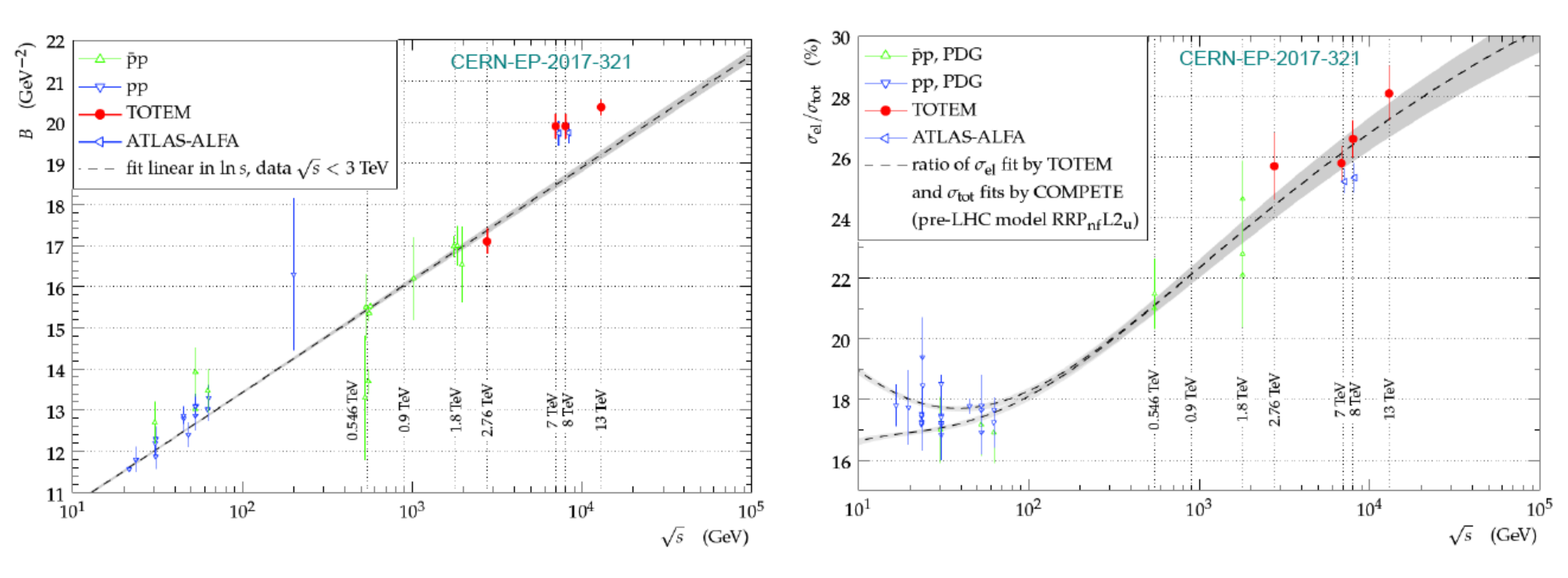}
\caption{$B$ slope as a function of $\sqrt{s}$ and increase of the elastic cross section as a function of energy.}
\label{bslope}
\end{figure}

The TOTEM collaboration also measured the differential elastic $pp$ $d \sigma/dt$ cross sections at center-of-mass energies of 2.76, 7, 8 and 13 TeV, that allows to measure the $B$ slope of the elastic cross section as a function of $\sqrt{s}$ as shown in Fig.~\ref{bslope}, left. The $B$ slope
is found to be larger  at 13 TeV and a
linear behavior (in $ln s$) of the elastic $d\sigma/dt$  cross section is only observed for $\sqrt{s}<3$ TeV, whereas it is no longer the case at higher energies~\cite{totemdata7,totemdata8,totemdata13}. The increase of $\sigma_{el}/\sigma_{tot}$ with energy is confirmed
at LHC energies, as shown in Fig.~\ref{bslope}, right.

A simple exponential fit to $d \sigma / dt$ at low $t$ fails to describe the elastic data at low $|t|$. The TOTEM collaboration performed an exponential fit of elastic data as $d \sigma/dt = A \exp (-B(t) |t|)$ where
different polynomial fits of $B(t)$ can be performed 
\begin{itemize}
\item $N_b=1$ $B=b_1$
\item $N_b=2$, $B=b_1+b_2t$
\item $N_b=3$, $B=b_1+b_2t+b_3 t^2$ 
\end{itemize}
where $N_b$ gives the number of parameters in the fit. 
A pure simple exponential form ($N_b=1$) is excluded at 7.2 $\sigma$ with 8 TeV data~\cite{totemdata8} and  similar results were found using using 13 TeV data~\cite{totemdata13}. Data cannot distinguish between a linear or a quartic dependence of $B$ on $|t|$.

\begin{figure}[h]
\centering
\includegraphics[width=0.7\textwidth]{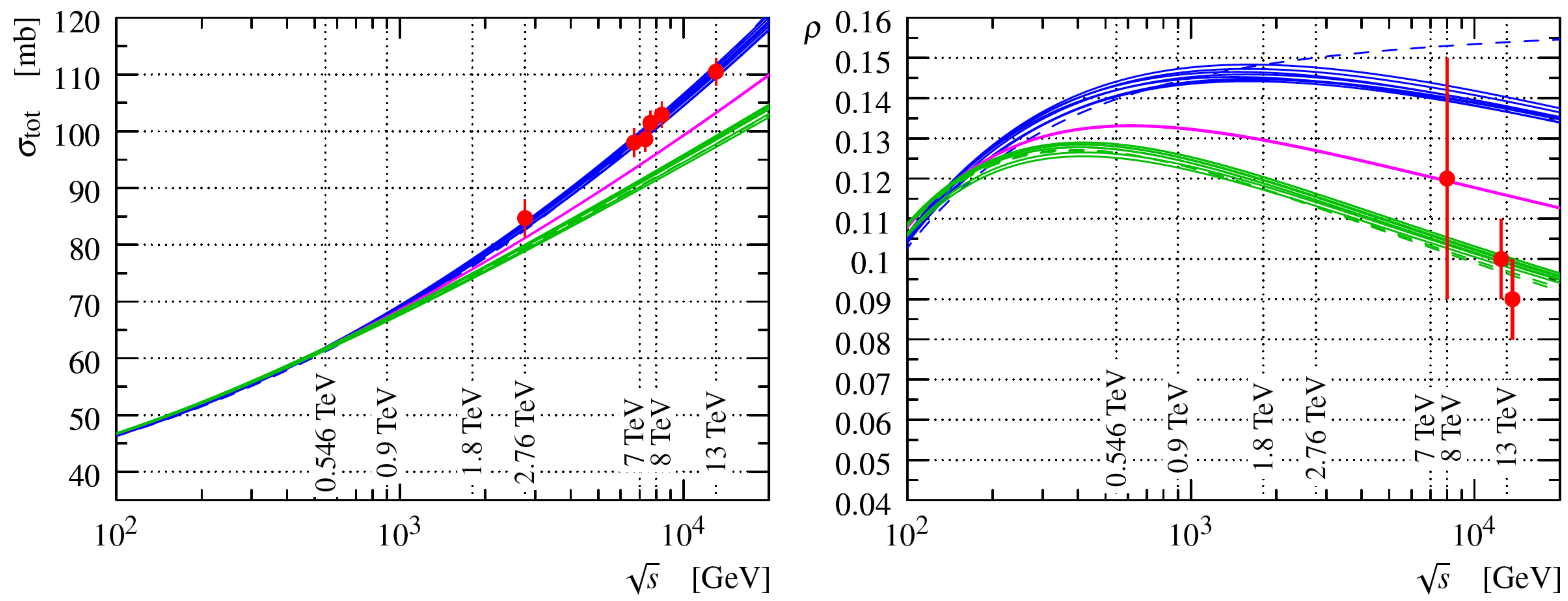}
\caption{Measurement of $\rho$ and $\sigma_{tot}$ from TOTEM as a function of $\sqrt{s}$.}
\label{rho}
\end{figure}

The very precise TOTEM elastic data  at 13 TeV and the fact that it was possible to use very high $\beta^*$ data (about 2 km) allowed to perform a precise measurement of elastic $pp$ $d\sigma/dt$ over a wide range in $|t|$ and to cover the Coulomb-Nuclear interference (CNI) region at $|t|\sim 10^{-4}-10^{-3}$ GeV$^2$ where $d\sigma/dt$ is sensitive to the amplitude in the Coulomb and nuclear regions
\begin{eqnarray}
\frac{d \sigma}{dt} \sim |A^C + A^N ( 1 -\alpha G(t))|^2. 
\end{eqnarray}
In the CNI region, both the modulus and the phase of the nuclear 
amplitude can be used to determine the $\rho$ parameter,
where the modulus is constrained by the measurement in the hadronic region and the
phase by the $t$ dependence.  The total cross section and the $\rho$ measurements performed by the TOTEM experiment are displayed in Fig.~\ref{rho} as a function of $\sqrt{s}$~\cite{totem2,rho}. The total cross section measurements are compared with different series of parametrizations from COMPETE~\cite{compete}, namely $\sigma_{tot}=a+b ln s + c ln^2 s$ in blue, $\sigma_{tot}=a+b ln^2 s $ in purple, and $\sigma_{tot}=a+b ln s $ in green. It is clear that $\sigma_{tot}$ favors the blue series of parametrizations whereas $\rho$ favors the green ones, especially at 13 TeV where  $\rho=0.09 \pm 0.01$. This creates a tension between these two measurements. It is worth noting that none of these parametrizations contains an odderon contribution, and adding the exchange of the Odderon in
addition to the Pomeron at high energies removes the tension between these measurements. This result can thus be interpreted as a first evidence for the odderon. For the models included in 
COMPETE and the Durham models~\cite{durham}, the TOTEM $\rho$ measurement at 13 TeV provides a 3.4 to 4.6$\sigma$ 
significance.

\section{The odderon discovery by the TOTEM and D0 collaborations}

In this section, we will describe the odderon discovery by the TOTEM and D0 collaborations~\cite{ourpaper,odderon1,odderon2}.
As we already mentioned in the previous section,  elastic $pp$ or $p \bar{p}$ scattering at high energies is due to the exchange of colorless objects, namely Pomeron and Odderon~\footnote{At lower energies such as the ISR, the situation is more complicated since there can be in addition exchanges of mesons and reggeons.}. 
Pomeron and Odderon~\cite{nicolescu,martynov,landshoff1,landshoff2,landshoff3,leader1,leader2,leader3} correspond respectively to positive and negative $C$ parity. From the QCD point of view, the pomeron is made of and even number of gluons (2, 4, 6, etc) which leads to a ($+1$) parity whereas the odderon is made of and odd number of  gluons (3, 5, 7, etc) corresponding to a ($-1$) parity.
The scattering amplitudes can be written as:
\begin{eqnarray}
A_{pp} &=& Even~+~Odd \\
A_{p \bar{p}} &=& Even~-~Odd. 
\end{eqnarray}
From the equations above, it is clear that observing a difference between $pp$ and $p \bar{p}$ interactions would be a clear way to observe the odderon, and this is the strategy followed by the TOTEM and D0 collaborations.

\begin{figure}[h]
\centering
\includegraphics[width=0.4\textwidth]{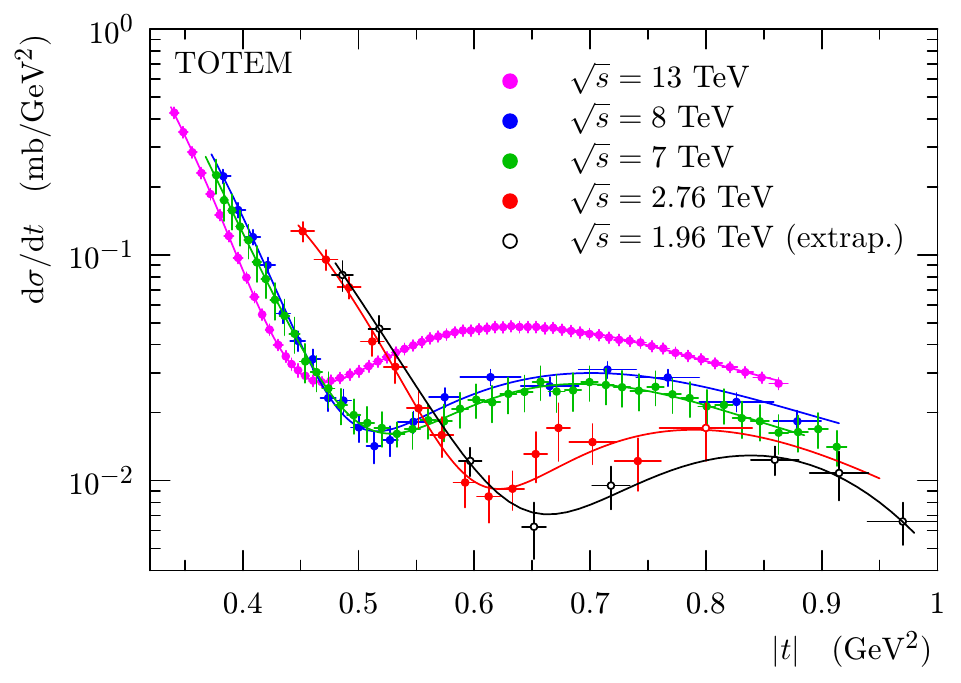}
\includegraphics[width=0.37\textwidth]{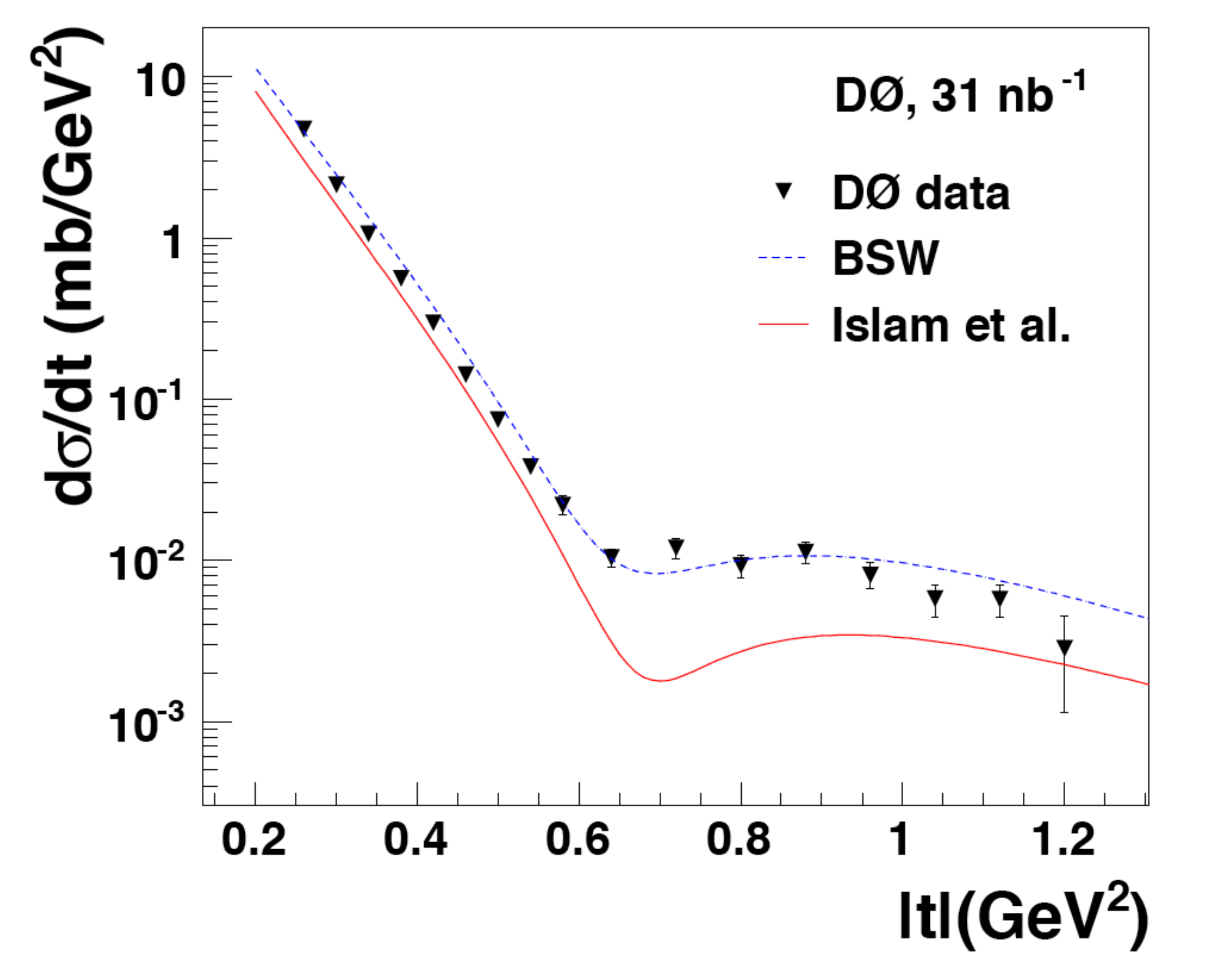}
\caption{Left: Comparison of elastic $pp$ $d\sigma/dt$ cross section measurements for different $\sqrt{s}$ by the TOTEM collaboration. Right: Elastic $p \bar{p}$ $d\sigma/dt$ cross section measurements from the D0 collaboration at 1.96 TeV.}
\label{data}
\end{figure}

\begin{figure}[h]
\centering
\includegraphics[width=0.5\textwidth]{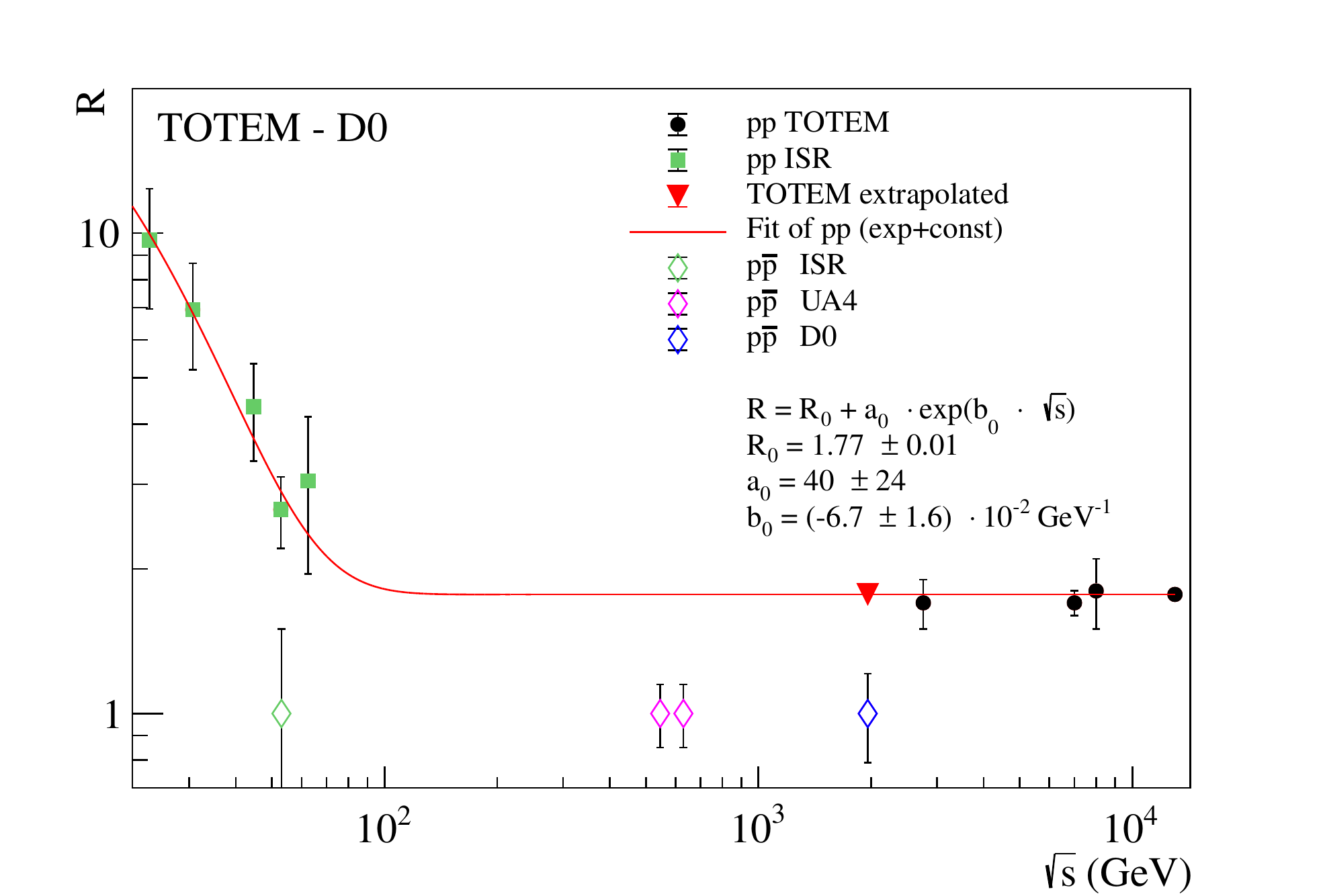}
\caption{Bump over dip ratio from the elastic $d \sigma/dt$ cross section measurements for $pp$ and $p \bar{p}$ interactions.}
\label{bumpoverdip}
\end{figure}

The elastic $pp$ $d\sigma/dt$ cross section measurements for different $\sqrt{s}$ by the TOTEM collaboration~\cite{totem,totemdata276,totemdata7,totemdata8,totemdata13} are shown in Fig.~\ref{data} together with the measurement for $p \bar{p}$ at the Tevatron by the D0 collaboration. $pp$ elastic cross sections always show the same features, namely a decreasing cross section at low $|t|$, the presence of a minimum, the ``dip" and of a maximum, the ``bump" at medium $|t|$ and finally a decreasing cross section at higher $|t|$. On the contrary, the $p \bar{p}$ cross section does not show any bump or dip. The first observable that shows a difference between $pp$ and $p \bar{p}$ elastic $d\sigma/dt$ cross sections is the ratio of the cross sections at the bump and at the dip as a function of $\sqrt{s}$~\cite{ourpaper,totem7} as displayed in Fig.~\ref{bumpoverdip} for LHC and Tevatron data~\cite{fpd,d0data} as well as from measurements from fixed targets or colliders at lower $\sqrt{s}$~\cite{isr1,isr2,isr3,isr4,isr5}. 
The ratio in $pp$ elastic collisions decreases as a function of $\sqrt{s}$ up to $\sim$ 100 GeV and is flat above. 
The D0 $p \bar{p}$  data show a ratio of 1.00$\pm$0.21 given the fact that no  bump nor dip is observed in the data within uncertainties. It leads to a difference of more than 3$\sigma$  between $pp$ and $p \bar{p}$ elastic data (assuming a flat behavior above $\sqrt{s} = 100 GeV$ for the $pp$ ratio). 

\begin{figure}
\centering
\includegraphics[width=0.99\textwidth]{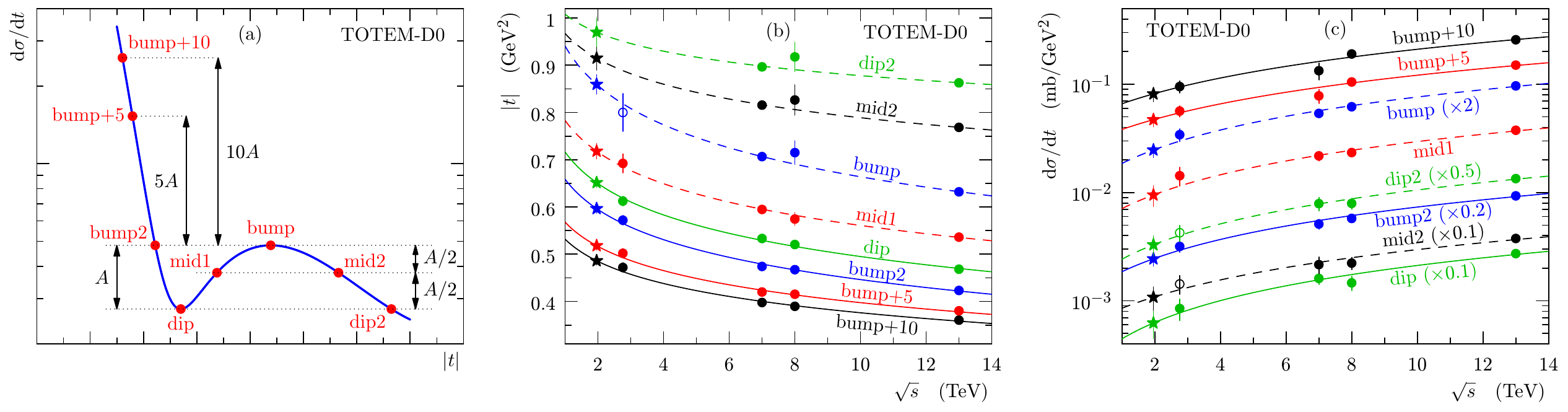}
\caption{$(a)$ Schematic definition of the characteristic points in the TOTEM differential cross section data. $(b)$ and $(c)$ Characteristic points in $(b)$ $|t|$ and $(c)$ $d\sigma/dt$ from TOTEM measurements at 2.76, 7, 8, and 13 TeV (circles) as a function of $\sqrt{s}$ extrapolated to Tevatron center-of-mass energy (stars). }
\label{characteristic}
\end{figure}

In order to get a better comparison of $pp$ and $p \bar{p}$ elastic data, it is needed to extrapolate the elastic $pp$ $d\sigma/dt$ 
data to 1.96 TeV, the center-of-mass energy of the Tevatron~\footnote{Unfortunately, running the LHC at 1.96 TeV, the Tevatron energy, is not easy and would lead to zero acceptance in the dip/bump region where we compare $pp$ and $p \bar{p}$ elastic data because of the roman pot detector acceptance.}. 
In order to predict the $pp$ elastic $d\sigma/dt$ cross section at 1.96 TeV, we define eight characteristic points of the 
behavior of the elastic cross section at 2.76, 7, 8 and 13 TeV and study the dependence of the $|t|$ and $d\sigma/dt$ 
values of these characteristic points as a function of $\sqrt{s}$. As we mentioned already, the $d \sigma/dt$ distributions for
 $pp$ collisions show a dip and a bump as a function of $|t|$ (see Fig.~\ref{data}). The chosen characteristic points are thus 
 the $dip$. the $bump$, $dip-2$ and $bump-2$ (same $d\sigma/dt$ as a the dip and the bump but at lower and higher $t|$),
$mid1$, $mid2$ (where the cross section is the average between the ones at the bump and at the dip), and 
$bump+5$ and $bump+10$ (where the difference between the cross sections is 5 or 10 times the differences between 
the ones at the bump and at the dip) as shown in Fig.~\ref{characteristic}, left.
This leads to  a distribution of $t$ and $d \sigma/dt$ values as a function of $\sqrt{s}$ for all characteristic points as shown in Fig.~\ref{characteristic}, middle and right~\cite{ourpaper}. We notice that the same $\sqrt{s}$ dependence can be used for all characteristic points for the $|t|$ and $d\sigma/dt$ values, namely 
\begin{eqnarray}
|t| &=& a \log (\sqrt{s}{\rm [TeV]}) + b  \label{eqn2} \\
(d\sigma/dt) &=& c \sqrt{s}~{\rm [TeV]} + d \label{eqn3}. 
\end{eqnarray}
From the fits, we determine the $|t|$ and $d\sigma/dt$ values of all characteristic points at the Tevatron $\sqrt{s}=1.96$ TeV as illustrated from the stars in Fig.~\ref{characteristic}.

The last step is to predict the $d \sigma/dt$ values of elastic $pp$ interactions in the same $|t|$ bins as the D0 measurement in order to make a direct comparison.
We fit the reference points extrapolated to 1.96 TeV from TOTEM measurements  using a double exponential fit ($\chi^2=0.63$ per dof)
\begin{eqnarray}
h(t) = a_1 e^{-b_1 |t|^2 - c_1|t|}  
+ d_1 e^{-f_1 |t|^3 - g_1 |t|^2 -h_1 |t|}.
\end{eqnarray}
This function is chosen for fitting purposes only.
The two exponential terms cross around the dip, one rapidly falling and becoming 
negligible in the high $t$-range where the other term rises above the dip.
In addition, differences in normalization taken into account by
adjusting TOTEM and D0 data sets to have the same cross sections at the optical point $d\sigma/dt(t = 0)$.

\begin{figure}[h]
\centering
\includegraphics[width=0.5\textwidth]{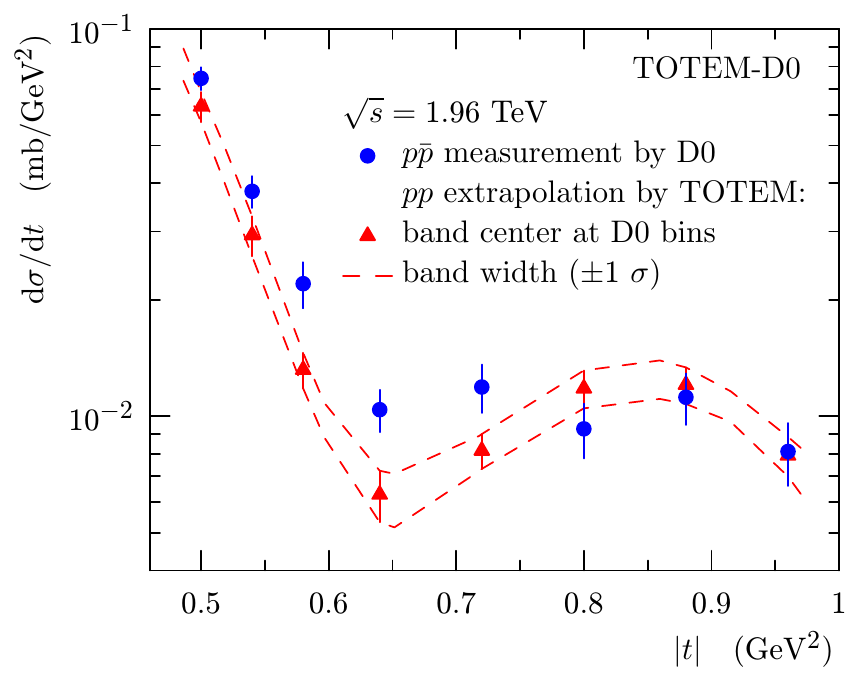}
\caption{Comparison between the D0 elastic $p \bar{p}$ $d \sigma/dt$ cross section measurement at 1.96 TeV and the extrapolated TOTEM measurement for $pp$ interactions}
\label{d0totem}
\end{figure}

Fig.~\ref{d0totem} provides the final result leading to the odderon discovery~\cite{ourpaper}. We display the comparison between the D0 $pp$ $d\sigma/dt$ and the extrapolated $pp$ $d\sigma/dt$ TOTEM measurements in the medium $|t|$ region of the dip and the bump. 
The $\chi^2$ test with six degrees of freedom yields the $p$-value of 0.00061, corresponding to a significance of 3.4$\sigma$. 

This result can be combined with the independent evidence of the 
odderon found by the TOTEM Collaboration using  $\rho$ and total cross section measurements at low $|t|$ at 13 TeV in a 
completely different kinematical domain described in the previous section. The combined 
significance ranges from  5.3 to  5.7$\sigma$ depending on the model. This constitutes the first experimental observation of the odderon, a major discovery at CERN and Tevatron.

\begin{figure}[h]
\centering
\includegraphics[width=0.4\textwidth]{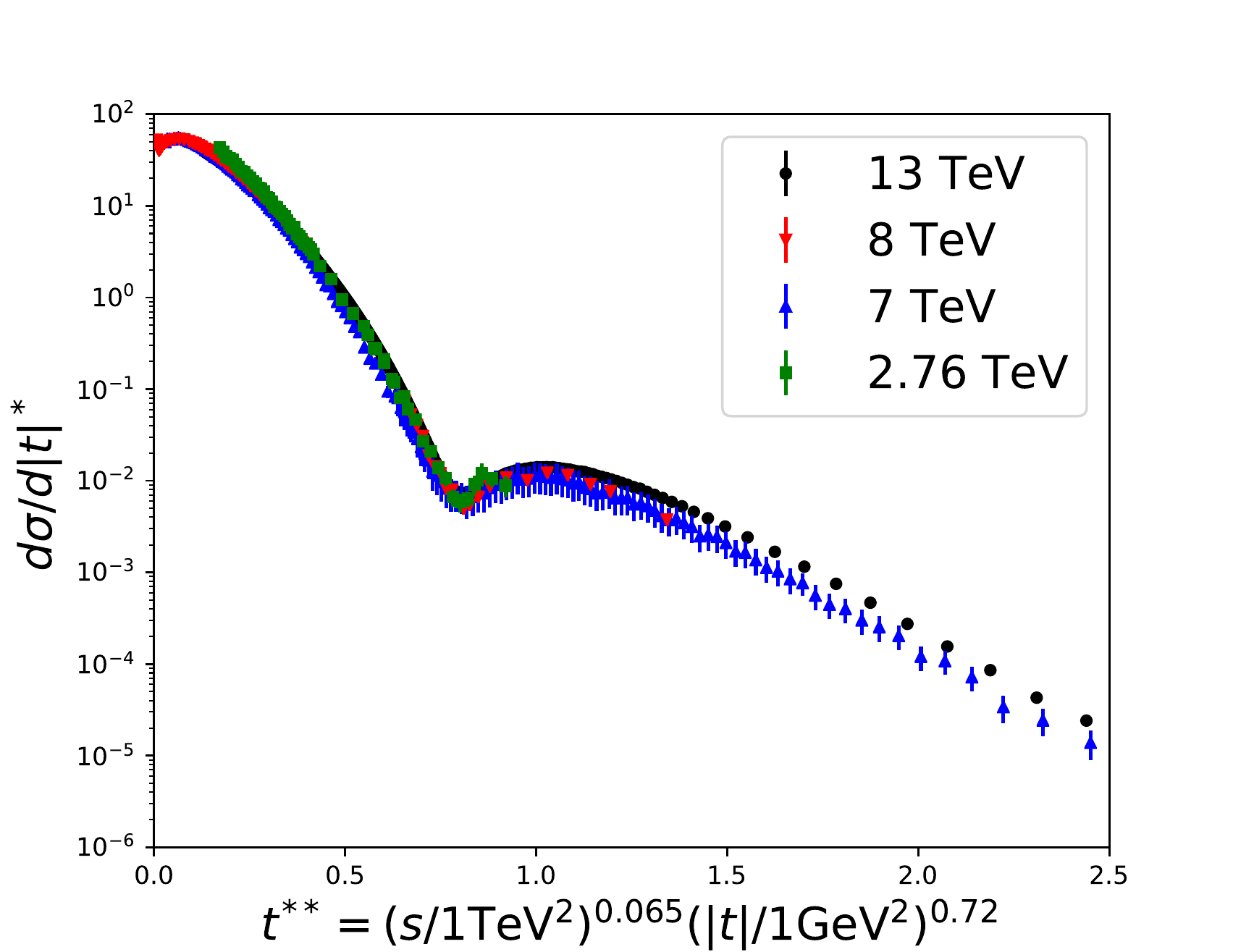}
\includegraphics[width=0.4\textwidth]{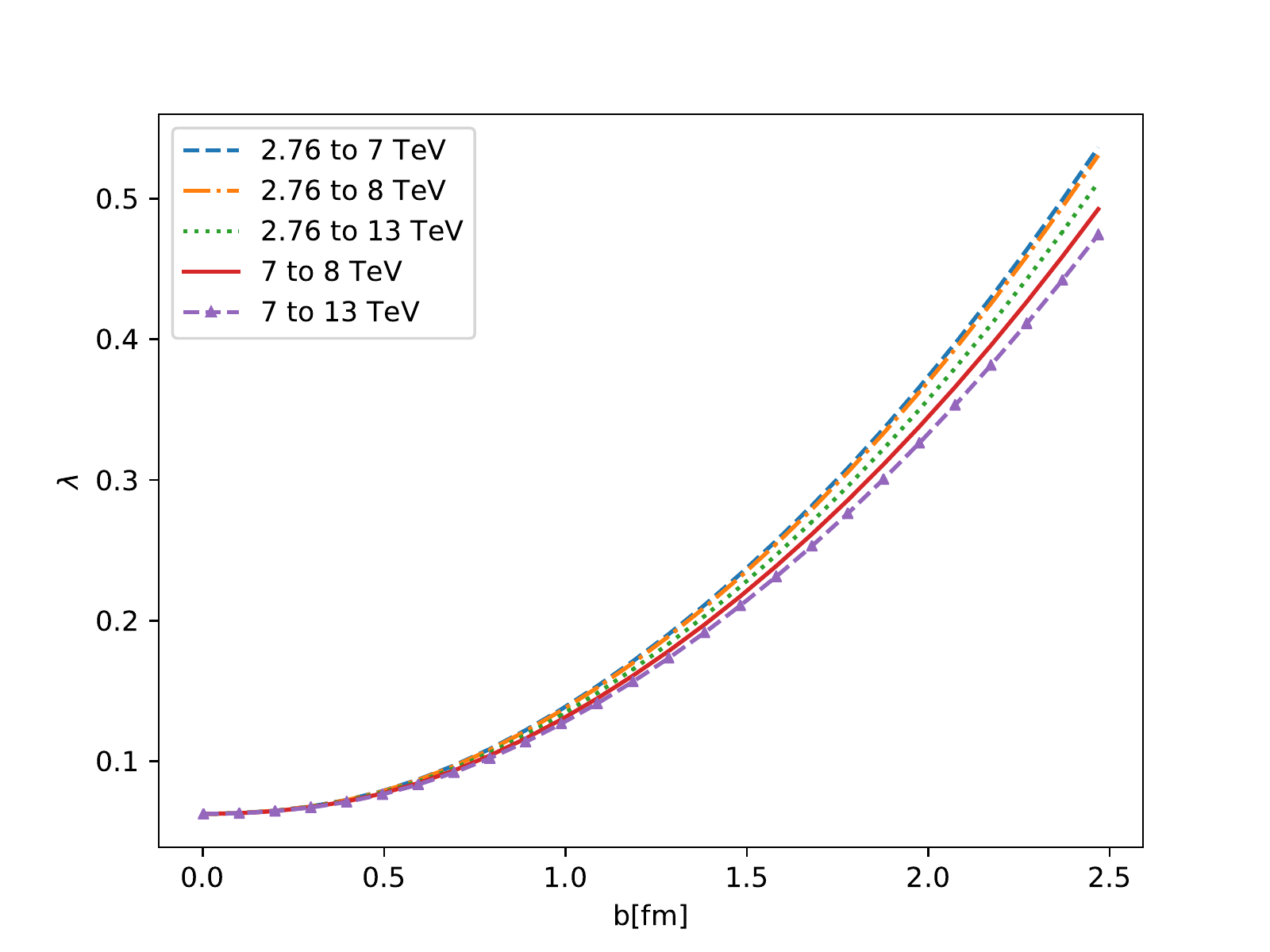}
\caption{Left: $\mathrm{d}\sigma/\mathrm{d}t^*$ as a function of $t^{**}$ showing the scaling of all TOTEM elastic scattering data at $\sqrt{s}=$ 2.76, 7, 8 and 13 TeV. Right: Power growth exponent $\lambda$ as a function of $b$ for various reference $\sqrt{s}$ pairs.}
\label{scaling}
\end{figure}

Let us discuss in addition the fact that the same $\sqrt{s}$ dependence can be used for all characteristic points. 
It indicates that there is a new scaling~\cite{scaling} in the elastic data measured by TOTEM as shown in Fig.~\ref{scaling}. If we plot 
$  \mathrm{d}\sigma/\mathrm{d}|t| \times s^{-0.305} $  as a function of $t^{**}=s^{0.065} (|t|)^{0.72}$, we see that all data for different $\sqrt{s}$ superimpose each other as shown on Fig.~\ref{scaling}, left.  In Fig.~\ref{scaling}, right, we display $\lambda$ as a function of $b$, the impact parameter, for various reference $\sqrt{s}$ pairs where $\lambda$ is defined as 
\begin{eqnarray}
 \lambda = \frac{1}{\ln (s_1/s_2)} \ln \Big(\frac{\mathrm{Re} \Gamma(s_1,b)}{\mathrm{Re} \Gamma(s_2,b)}\Big) \; .
\end{eqnarray}
where $s_1$ and $s_2$ are two center-of-mass energies square, and the $\Gamma$s are the profile functions in the $b$-parameter space. $\lambda$ is equal to 0.06 at small values of $b$ and goes up to 0.4 at large values of $b = 2$ fm. The power exponent $\lambda$ has a weak dependence on the reference energies $\sqrt{s}$ used to extract it. In particular, at $b = 0$, the value of $\lambda = 0.06$ does not depend on the reference energies and is directly related to the scaling properties. The $\lambda$ values at small $b$ are compatible with expectations from a dense object, such as a black
disc~\cite{scaling}.

\section{Conclusion}
In this short report, we first described the total, inelastic and elastic cross sections measured by the TOTEM collaboration at 2.76, 7, 8 and 13 TeV, that led to a non-exponential behavior of $d\sigma/dt$. 
A detailed comparison between $p\bar{p}$ (1.96 TeV from D0) and $pp$ (2.76, 7, 8, 13 TeV from TOTEM) elastic $d\sigma/dt$ data led to the fact that $pp$ and $p\bar{p}$ cross sections differ
with a significance of 3.4$\sigma$ in a model-independent way and thus provides evidence that the Colorless $C$-odd gluonic compound  i.e. the odderon
is needed to explain elastic scattering at high energies. 
When combined with the $\rho$ and total cross section result at 13 TeV, the significance is in the range 5.3 to 5.7$\sigma$ and thus constitutes the first experimental observation of the odderon, a major discovery at CERN and Tevatron.

Many additional recent physics results of the TOTEM and CMS collaborations include the sensitivity to quartic anomalous couplings, benefitting from the fact that the LHC can be considered as a $\gamma \gamma$ collider, quasi-real photons being emitted by the protons. This leads to unprecedented sensitivities to quartic anomalous couplings such as $\gamma \gamma \gamma \gamma$~\cite{gammagamma,gammagamma2,gammagamma3, gammagamma4,gammagamma5,gammagamma6,gammagamma7,gammagamma8} $\gamma \gamma WW$~\cite{ww,ww1}, $\gamma \gamma ZZ$, $\gamma \gamma \gamma Z$~\cite{gammaz} and $\gamma \gamma t  \bar{t}$~\cite{ttbar,ttbar1} and to axion-like particles at high mass~\cite{axion,axion1} with reach usually better by two or three orders of magnitude with respect to standard methods at the LHC without tagging the intact protons in the final state.





\nolinenumbers

\end{document}